\documentclass[nofootinbib,preprintnumbers,amsmath,amssymb, 11pt]{article}
\usepackage[english]{babel}
\usepackage[a4paper, inner=2cm, outer=2cm, top=3cm, bottom=3cm]{geometry}
\usepackage[dvipsnames]{xcolor}
\usepackage{mathtools}
\usepackage{pstricks}
\usepackage{caption}
\usepackage{graphicx}
\usepackage{amsmath}
\usepackage{amssymb}
\usepackage{mathcomp}
\usepackage{cancel}
\usepackage{textcomp}
\usepackage{enumitem}
\usepackage{physics}
\usepackage{romannum}
\usepackage[doublespacing]{setspace}
\usepackage{subcaption}
\usepackage[linktoc=all]{hyperref}
\usepackage{booktabs}
\usepackage{cite}
\usepackage[utf8]{inputenc}
\usepackage{authblk}
\usepackage{bm}

\usepackage{array}
\usepackage{graphicx}
\usepackage{caption}
\usepackage{lipsum}
\usepackage{empheq}
\usepackage[many]{tcolorbox}
\usepackage[normalem]{ulem}

\definecolor{DarkBlue}{RGB}{0,38,77}
\definecolor{PersianBlue}{RGB}{28,57,187}

\hypersetup{
  colorlinks   = true,
  urlcolor     = magenta, 
  linkcolor    = PersianBlue,
  citecolor    = Red
}

\makeatletter
\renewcommand{\eqref}[1]{%
  \textup{\hyperref[#1]{\textcolor{PersianBlue}{(\ref*{#1})}}}%
}
\makeatother

\tcbset{highlight math style={enhanced,
colframe=black,colback=white,arc=2mm,boxrule=0.9pt}}

\newcommand{\calO}{{\cal{O}}}
\newcommand{\calV}{{\mathcal{V}}_{\text{int}}}
\newcommand{\calW}{{\mathcal{W}}_{\text{int}}}
\newcommand{\nc}{\newcommand}
\nc{\ba}{\begin{eqnarray}}
\nc{\ea}{\end{eqnarray}}
\newcommand\be{\begin{equation}}
\newcommand\ee{\end{equation}}
\newcommand\Mpl{{M_{\rm Pl}}}

\title{\bf Non-relativistic effective theories for fields with general potentials and their implications for cosmology}
\author[1, 2]{H. S. Modirzadeh\thanks{\href{mailto:hsmodir@ipm.ir}{\ hsmodir@ipm.ir}}}
\author[2]{R. Moti\thanks{\href{mailto:r.moti@ipm.ir}{\ r.moti@ipm.ir}}}
\author[2]{M. H. Namjoo\thanks{\href{mailto:mh.namjoo@ipm.ir}{\ mh.namjoo@ipm.ir}}}
\affil[1]{{\small\textit{Department of Physics, Sharif University of Technology, Tehran, Iran}}}
\affil[2]{{\small\textit{School of Astronomy, Institute for Research in Fundamental Sciences (IPM), \hspace{4cm} Tehran, Iran P.O. Box 19395-5531}} }
\date{}

\begin{document}
\maketitle
\pagenumbering{arabic}

\begin{abstract}

Non-relativistic effective field theories (NREFTs) play a crucial role in various areas of physics, from cold atom experiments to cosmology.  In this paper, we present a systematic framework for deriving NREFTs from relativistic theories with generic self-interactions. Our approach allows for (but is not limited to) non-power-law potentials (such as those arising from dilatons or axions) or potentials that are non-analytic around the classical vacuum (such as those with logarithmic radiative corrections). These are of theoretical and phenomenological interest but have largely been unexplored in the non-relativistic regime. NREFTs are typically viewed as approximations for systems with low velocities, weak couplings, and small field amplitudes. The latter assumption is relaxed in our approach, as long as the mass term remains dominant (ensuring the validity of the non-relativistic limit). Additionally, we establish an effective fluid description for the non-relativistic scalar field, identifying key quantities such as energy density, pressure, and sound speed. To enable cosmological applications, we extend our formalism to account for the expanding universe, providing a reliable tool for investigating ultra-light dark matter models with arbitrary self-interactions. Finally, we demonstrate the applicability of our NREFT in analyzing solitons, which is also relevant to cosmology for studying celestial objects such as boson stars and the cores of dark matter halos. 

\end{abstract}

\newpage

\tableofcontents

\section{Introduction}
\label{Sec:intro}
The non-relativistic limit of a (possibly ultra-violate) theory is a useful tool for understanding low energy and low velocity phenomena. 
Successful examples of the non-relativistic limit arise in various fields of physics. The dynamics of heavy quarks at low energies are described by non-relativistic quantum chromodynamics (QCD), an effective field theory (EFT) that provides a controlled approximation to the full relativistic QCD Lagrangian, by systematically integrating out high energy degrees of freedom \cite{Lepage:1988, Pineda:1998}.
Another prominent example is found in cold atom experiments, where neutral atoms are cooled to ultra-cold temperatures (often in the nano-Kelvin range), making non-relativistic descriptions highly accurate \cite{Pethick:2008, Pitaevskii:2016,Deng:2018xsk}. Since the kinetic energy of atoms is extremely small compared to their rest energy, relativistic effects become negligible, and the system is consistently described by the non-relativistic Schr\"odinger-like field equations rather than the relativistic Dirac or Klein-Gordon equations.

In cosmology, non-relativistic effective theories are employed to describe the dark matter systems. If dark matter particles are very light --- which is the case, e.g., for axion-like or ultra-light dark matter --- the occupation number is very high, allowing for the application of the mean field approximation and describing the system as a classical field \cite{Guth:2014hsa}. In this regime, dark matter may form a coherent condensate state. At the same time, since dark matter must be cold, one can take the non-relativistic limit and obtain a classical non-relativistic effective field theory (NREFT) for the  condensate of the dark matter particles \cite{Namjoo:2017nia}. The resulting effective theory is also suitable for studying compact objects such as boson stars and solitons, as long as they remain non-relativistic \cite{Braaten:2018lmj, Mukaida:2016hwd, Chavanis:2011,Chavanis:2017loo, Visinelli:2017ooc}.

Describing non-relativistic systems as fluids is also of interest in various fields, in particular, in cosmology \cite{Weinberg:2008, Mukhanov:2005, Salehian:2020bon}. Such an alternative framework is particularly suitable when general relativistic effects --- such as the expansion of the universe --- become prominent (even though the field of interest remains non-relativistic). The fluid description enables the convenient use of Einstein's equations, with the fluid's energy-momentum tensor on the right-hand side. 

Unlike standard scalar field theories with a possibly power-law self-interaction, more general self-interacting scalar field models in the non-relativistic limit have received comparatively less attention.  However, non-power-law interactions such as those found in the Coleman-Weinberg model \cite{Coleman:1973, Meissner:2007}, or in the dilaton \cite{Bardeen:1986, Gasperini:2007}, or axion theories \cite{axions} arise naturally in theoretical physics, making them particularly compelling for further study. Therefore, it is well-motivated to develop a framework that can be applied to such theories to obtain an NREFT that describes their corresponding low energy dynamics.

In this work, we focus on such models, developing an EFT for a massive scalar field with a very general potential in the low energy limit.  This is in contrast with the EFTs presented in Refs.~\cite{Namjoo:2017nia}, \cite{Salehian:2020bon}, and \cite{Salehian:2021khb} that are restricted to the power-law self-interactions. Here, the potential is allowed to be non-power-law (such as those from the dilaton-like theories) or non-analytic around the classical vacuum (such as the Coleman-Weinberg-like potentials). The field amplitude is also allowed to be large, as long as the mass term remains dominant. 
We begin with an analysis in Minkowski spacetime, but then obtain the leading effective theory in an expanding universe, taking into account up to linear perturbations around the isotropic and homogeneous universe. We also re-express our results in an effective fluid description and demonstrate how fluid quantities, such as the equation of state and sound speed, are influenced by a generic potential.

The resulting EFT is also applicable for studying non-relativistic solitonic solutions. We provide the leading order equations for that purpose and, as an application, investigate the profile of solitons in the presence of complex potentials. Additionally, we discuss how an energy balance analysis can be performed in this context and for a relatively generic profile. This may be used, e.g., to study the mass-radius relation of solitons formed from a generic self-interacting force that balances other forces.

In the next section, we introduce an analytic procedure for deriving the NREFT of a classical scalar field with a general potential. Once we establish the leading order NREFT, Sec.~\ref{SecIII} recasts the theory in an effective fluid description. Sec.~\ref{Sec:EFT_FRW} then extends the results to account for the expansion of the universe. Building on this foundation, in Sec.~\ref{Sec:examples} we explore several noteworthy examples of self-interacting models. In Sec.~\ref{SecIV}, we present the effective equations for studying solitonic solutions, use numerical methods to examine the profile of solitons in a few explicit examples, and provide the equations required for the energy balance analysis. Finally, Sec.~\ref{Sec:conclusion} summarizes our findings.

\section{Non-relativistic effective field theory}
\label{SecII}

In this section, we present an analytic method for deriving the NREFT from a theory of a relativistic scalar field with a general potential. This method generalizes the framework developed in Ref.~\cite{Namjoo:2017nia}. In this paper, we focus on the leading order EFT. For this purpose, it is sufficient to work in Minkowski spacetime. Once we obtain the EFT in this limit, incorporating additional contributions in a (linearly) perturbed Friedmann--Lemaître--Robertson--Walker (FLRW) spacetime will be straightforward by adapting the results of Ref.~\cite{Salehian:2020bon}, which will be discussed in Sec.~\ref{Sec:EFT_FRW}. 
We thus consider the following Lagrangian density of the real scalar field $\phi \equiv \phi(t,\vb{x})$
\begin{equation}
\label{Lphi}
\mathcal{L}_{\phi} = - \dfrac{1}{2} \eta^{\mu\nu} \partial_{\mu}\phi \partial_{\nu}\phi - \dfrac{1}{2} m^2 \phi^2 - V_{\text{int}} (\phi) \, ,
\end{equation}
where we used the mostly positive metric signature and, to have a consistent non-relativistic limit, we assume that the mass term dominates over the self-interaction potential.  

The standard procedure for the EFT derivation in the non-relativistic limit starts with a  field redefinition as follows
\begin{equation}
\label{redef}
 \phi(t,\vb{x}) = \dfrac{1}{\sqrt{2m}} \left( \psi(t,\vb{x}) e^{-imt} +\psi^*(t,\vb{x}) e^{imt} \right)\, .
\end{equation}
The complex scalar field $\psi(t,\vb{x})$ --- which we will henceforth refer to as the ``non-relativistic field" --- is assumed to vary slowly in time compared to $\phi(t,\vb{x})$ which rapidly oscillates with frequency $m$. This redefinition is intended to separate the slow and fast dynamics of $\phi(t,\vb{x})$. The redefinition according to Eq.~\eqref{redef} ensures that $\abs{\psi(t,\vb{x})}^2$ approximately corresponds to the number density of particles, making $\psi(t,\vb{x})$ a convenient variable for analyzing the system in the non-relativistic regime.

To promote the field redefinition of Eq.~\eqref{redef} to a consistent canonical transformation, we remove the  redundancy by also redefining the conjugate momentum and consider the following transformations
\begin{align}
\label{full-redef}
&\phi = \dfrac{1}{\sqrt{2m}} \left( \psi e^{-imt} +\psi^* e^{imt} \right) \, ,
&\dot \phi = -i \sqrt{\dfrac{m}{2}} \left( \psi e^{-imt} - \psi^* e^{imt} \right) \, .
\end{align}
This transformation may be implemented in the Lagrangian by introducing a Lagrange multiplier that enforces the constraint $\dot \psi \, e^{-imt}+\dot \psi^* e^{imt}=0$, which is implied by the transformation. We refer the readers to Ref.~\cite{Salehian:2020bon} for further details and here only present the final form of the Lagrangian for the non-relativistic field, after implementing the field transformation, adding the Lagrange multiplier, and then integrating out non-dynamical variables:
\begin{equation}
\mathcal{L}_{\psi,\psi^*} = \dfrac{i}{2}(\psi^*\dot{\psi}-\psi\dot{\psi^*})- \dfrac{1}{2m} \grad\psi .\grad\psi^*  - \bigg\{ \dfrac{ e^{2imt}}{2m}   \grad\psi . \grad\psi+ \text{c.c.} \bigg\}
- {V}_{\text{int}}(\psi,\psi^*)\, ,
\label{Lag-psi}
\end{equation}
where ${V}_{\text{int}}(\psi,\psi^*)$  is the self-interaction potential according to the original Lagrangian Eq.~\eqref{Lphi} with $\phi$ replaced by $\psi$ and $\psi^*$ according to the redefinition Eq.~\eqref{redef}. 
The resulting Schr\"odinger-like equation for $\psi$ is
\ba 
\label{psi-eom}
i \dot{\psi} =-\dfrac{1}{2m}\laplacian{\psi}  -\dfrac{e^{2imt}}{2m}\laplacian{\psi}^* + V_{\text{int},\psi^*  } \, ,
\ea 
where $V_{\text{int},\psi^*}$ denotes the derivative of the potential with respect to $\psi^*$. 
Note that the consistent transformation according to Eq.~\eqref{full-redef} automatically removes second order time-derivatives from the equation of motion. In fact, 
so far, there is no approximation and the field redefinitions and the resulting equations are exact. Now, one may take the non-relativistic limit of the theory and obtain an EFT that may be truncated at an arbitrary order. Generally, the EFT will be organized in powers of small quantities/operators in the non-relativistic limit which, schematically, may be expressed by
\ba 
\label{epsilons}
\epsilon_t \sim \dfrac{\partial_t}{m} \, , \qquad   \epsilon_x \sim \dfrac{\nabla^2}{m^2} \, , \qquad \epsilon_V \sim \dfrac{V_{\text{int}}}{m |\psi|^2} \, .
\ea 
The validity of the NREFT requires the smallness of all above operators when acted on the non-relativistic field.  In the expanding background, another small quantity appears in the EFT which may be described by $\epsilon_H \sim H/m$ where $H$ is the Hubble parameter.  A systematic derivation of the EFT as a power series in the above small parameters is presented in Refs.~\cite{Namjoo:2017nia,Salehian:2021khb,Salehian:2020bon}. Here, we work instead with the leading order EFT but consider a general potential and large field amplitudes.

In a low energy experiment, the system is probed at time scales much longer than $1/m$. To proceed with deriving the  EFT, we thus need to integrate out rapidly oscillating modes. This can be achieved by coarse-graining the equation of motion in time, following the approach developed in Refs.~\cite{Namjoo:2017nia, Salehian:2020bon}. This method is based on introducing a time-smearing (averaging) operation, which acts on an arbitrary time-dependent quantity $X(t)$ as
\begin{equation}
	\label{cg}
\expval{X(t)} = \int_{-\infty}^{\infty} \dd{t'} X(t') W(t-t') \ .
\end{equation}
The weighting window function $W(t-t')$ is chosen such that the smearing operation isolates the slowly varying components of $X(t)$, namely those localized around zero frequency. In Fourier space, this is achieved by adopting a top-hat window function,
\begin{equation}
\hat{W}(\omega)= 
\left\{
\begin{array}{ll}
    1  \qquad  \abs{\omega} < m/2\\
    0   \qquad  \text{otherwise}
\end{array}
\right.
\end{equation}
which removes frequency components separated from $\omega =0 $ by more than $m/2$. The corresponding window function in the time domain is then given by
\begin{equation}
W(t) = \int_{-\infty}^{\infty} \frac{\dd{\omega}}{2\pi}  \hat{W}(\omega) e^{i\omega t} = \dfrac{\sin(mt/2)}{\pi t}  \ .
\end{equation}
With this choice, any time-dependent quantity admits an exact mode decomposition as follows
\begin{equation}
X(t) = \sum_{\nu = -\infty}^{\infty} X_{\nu}(t) e^{i\nu m t} \ ,
\end{equation}
where the mode coefficients are defined by
\begin{equation}
 X_{\nu}(t) = \expval{X(t) e^{-i\nu m t}} = \int_{-m/2}^{m/2} \frac{\dd{\omega}}{2\pi}  \hat{X}(\omega + \nu m) e^{i\omega t} \ .
\end{equation}
This procedure eliminates highly oscillatory contributions while preserving their impact on the slowly varying dynamics. We  have the following useful identities as a result of the defined smearing operator:
\begin{equation}
	(X Y)_{\nu}=\sum_{\alpha} X_{\alpha}Y_{\nu-\alpha}, \qquad (X^*)_{\nu} = X^*_{-\nu}\equiv (X_{-\nu})^*, \qquad (X e^{i\alpha m t})_{\nu}=X_{\nu-\alpha}\, .
\end{equation}
Applying this formalism to the non-relativistic scalar field $\psi$, we note that although $\psi$ is assumed to be dominated by a slow mode in the nonrelativistic limit, nonlinear interactions generically induce subdominant, rapidly oscillating components. These fast-modes can backreact on the slow-mode and influence its long term behavior. While, at leading order in the EFT (which is our primary focus in this paper), we can neglect the fast-modes, we will see this in Sec.~\ref{SecIII} that a few fast-modes will contribute to the leading order effective fluid description.

Accordingly, we decompose the field $\psi$ into a slow mode and an infinite sum of fast modes as
\begin{equation}
	\psi =\sum_{\nu = -\infty}^{\infty} \psi_{\nu} \, e^{i\nu m t} 
	=
	{\psi}_s + \sideset{}{'}\sum_{\nu = -\infty}^{\infty} \psi_{\nu} \, e^{i\nu m t} \, ,
	\label{mode exp}
\end{equation}
where the prime on the summation indicates that the slow-mode ($\nu=0$) is excluded from the sum. Since the slow-mode is the mode of interest in the non-relativistic limit, it deserves a distinct notation and we denote it by $\psi_s$. The validity and generality of the above decomposition in the non-relativistic limit requires  $\psi_{\nu}$ to be slowly varying functions of time (rather than being constant). 
 Moreover, the non-relativistic limit is valid as long as $|\psi_\nu| \ll |\psi_s|$ for all $\nu \neq 0$. 

 The equation of motion for the mode $\nu$ may be obtained by substituting any function of time in Eq.~\eqref{psi-eom}  (such as $\psi$ and the potential) with their mode decomposition, multiplying it  by a factor of $e^{-i \nu m t}$  and coarse-graining the entire equation. The latter procedure eliminates all terms in the equation of motion that oscillate with a frequency of $m$ or higher (since these terms average to zero over a long period of time). This results in
\begin{equation}
	i (\dot{\psi}_{\nu} + i \nu m \psi_{\nu}) = -\dfrac{1}{2m} \nabla^2 \psi_{\nu} -\dfrac{1}{2m} \nabla^2 \psi^{*}_{2- \nu} +   (V_{\text{int},\psi^*  }  )_{\nu }\, ,
	\label{psinu-eom}
\end{equation}
where $(V_{\text{int},\psi^*  })_{\nu }$ is the mode $\nu$ of  $V_{\text{int},\psi^*}$. 
 In particular, for the slow-mode we have
\begin{equation}
	i \dot{\psi}_{s}   = -\dfrac{1}{2m} \nabla^2 \psi_{s} -\dfrac{1}{2m} \nabla^2 \psi^{*}_{2} +   (V_{\text{int},\psi^*  }  )_{s }\, .
	\label{psis-eom}
\end{equation}
We note that, even without self-interaction, the fast-modes affect the dynamics of the slow-mode. The self-interaction term adds extra contributions from the fast-modes. To obtain an effective equation of motion solely involving the slow-mode, we can solve Eq.~\eqref{psinu-eom} iteratively and substitute the solutions into Eq.~\eqref{psis-eom}. However, at leading order, we can disregard all the fast-modes (since $|\psi_\nu| \ll |\psi_s|$ for all $\nu \neq 0$). This does not render our EFT trivial, as the self-interaction term will take a different form from what it appears in the original theory, Eq.~\eqref{Lphi}, which we will explore in the next subsection.

\subsection{Leading effective potential}
\label{Sec-Veff}
 In this section, we obtain the leading order contribution to the equation of motion for the slow-mode from the self-interaction. This may be achieved by coarse-graining the original potential. That is, we compute 
 \ba 
 \label{calV_leading}
 {\mathcal{V}}_{\text{int}} \equiv \langle V_{\text{int}} (\psi_s,\psi_s^*) \rangle \, ,
 \ea 
 where $\langle \cdot \rangle$ denotes time coarse-graining (see Eq.~\eqref{cg}) and in the potential $V_{\text{int}}$ we replace $\phi$ with the right-hand side of Eq.~\eqref{redef} and replaced $\psi$ with $\psi_s$ to obtain the leading order contributions.  We work with the potential $V_{\text{int}}$ despite the fact that $V_{\text{int},\psi^*  }$  appears in the equation of motion Eq.~\eqref{psi-eom}. We will see that deriving $\calV$ suffices for obtaining  the leading order EFT. A similar method can be employed to extract fast-modes of the potential by multiplying appropriate factors of $e^{\pm imt}$ before coarse-graining. Likewise, a similar procedure applies to any function of the original field $\phi$, including derivatives of the potential.   
 
 We consider a very general potential and, depending on the analyticity properties of it, we adopt two methods to obtain the effective, coarse-grained potential. Our analytic method leads to a power series as the effective potential. Therefore, it is useful in situations where either the power series can be truncated or can be resummed. If neither is suited in a situation, Eq.~\eqref{calV_leading} may be utilized to obtain the coarse-grained potential numerically. 

\subsubsection{Analytic potentials}
\label{Sec_anal}
If the potential is analytic around the classical vacuum, $\phi=0$, the effective potential may be obtained by expanding the potential in power series and coarse-graining each term. More explicitly, we may write the potential in the following form
\begin{equation}
\label{expand_analytic}
V_{\text{int}}(\phi) = \sum_{n=0}^\infty \alpha_{n} \phi^{n}\, ,
\end{equation}
where $\alpha_{n}=\frac{1}{n!}\frac{\partial^n V_{\text{int}}}{\partial \phi^n}|_{\phi=0}$ represent the expansion coefficients. After the field redefinition according to Eq.~\eqref{redef}  and expanding the resulting binomial factor, each term   in Eq.~\eqref{expand_analytic} contains another expansion in powers of the non-relativistic field as follows
\begin{equation}
V_{\text{int}}(\phi) = \sum_{n=0}^\infty \sum_{k=0}^{n} \binom{n}{k} \dfrac{\alpha_{n} }{(\sqrt{2m})^{n}} \psi^k \psi^{* \ n-k} e^{-imkt+im(n-k)t} \, .
\end{equation}
Since here we are interested in the leading-order contribution, as described in Sec.~\ref{Sec-Veff}, we replace $\{\psi, \psi^*\}$ with $\{\psi_s, \psi_s^*\}$. Coarse-graining removes all oscillatory terms, leaving only terms for which the condition $n-2k=0$ is satisfied. Thus, a term survives after coarse-graining only if $n$ is even and if $k=n/2$, resulting in

\begin{equation}
	\label{calV_anal}
{\mathcal{V}}_{\text{int}} = \sum_{n=0}^\infty \alpha_{2n}  \binom{2n}{n} \dfrac{1}{({2m})^{n}} \abs{\psi_s}^{2n} \ .
\end{equation}

Note that, to the leading order, only terms with even powers of $\phi$ remain after coarse-graining because the odd terms always contain oscillatory factors. 
Note also that we do not necessarily require the power series Eq.~\eqref{expand_analytic} to converge. In fact, our method is most interesting in the large field limit where the series expansion cannot be truncated. In this case, a compact form may be obtained through resummation. See  Sec.~\ref{Sec:examples} for a few examples.

\subsubsection{Non-analytic potentials}
\label{Sec_non-anal}
If the potential is non-analytic around $\phi=0$, the expansion of the form of Eq.~\eqref{expand_analytic} is forbidden. However, it is still possible to  obtain the effective potential by expanding around a different point in the field-space. Examples of this situation include potentials with logarithmic factors (e.g., due to radiative corrections) or potentials like $\phi^{\beta}$ with $\beta$ a real but  non-integer constant. Since we are interested in the situations where a condensate of particles is formed, it is natural to expand around such a condensation where the potential is likely to be analytic.\footnote{We are not aware of any physical situation where the potential is non-analytic around the condensate and exclude it from our analysis.} To ensure that the potential remains real throughout the entire evolution, we assume that the potential is a function of $\phi^2$. This implies that, for example, $\phi^{\beta}$ must be understood as $(\phi^2)^{\beta/2}$. According to Eq.~\eqref{redef}, we have $\phi^2 = \frac{|\psi|^2}{m} \left(1+ Y\right)$ where $Y=\frac{1}{2z}+\frac{z}{2}$ with $z=\frac{\psi}{\psi^*}e^{-2imt}$. Note that $V_{\text{int}}(\phi^2)$ is a real function and $Y$ is a real parameter. Therefore, assuming that the potential is analytic around $Y=0$, we may express it in a power series as follows
\ba 
\label{Vexpand-nonanalytic}
V_{\text{int}}(\phi^2) = \sum_{n=0}^{\infty} \tilde \alpha_n(|\psi|^2)  \left( \frac{Y}{m} \right)^{n} |\psi|^{2n} \, ,
\ea 
where $\tilde \alpha_{n}=\frac{1}{n!}\frac{\partial^n V_{\text{int}}}{\partial \phi^{2n}}|_{\phi^2=|\psi|^2/m}$ (which, note that, is field-dependent). At leading order, the oscillatory factors appear in $Y$ (not in $\psi$). Once again, to leading order,  time averaging eliminates all terms that contain oscillatory factors leading to
\ba 
\label{V_expand_analytic}
 {\mathcal{V}}_{\text{int}} = \sum_{n=0}^{\infty} \tilde \alpha_{2n} (|\psi_s|^2)  \binom{2n}{n}    \dfrac{1}{(2m)^{2n}} \abs{\psi_s}^{4n} \, .
\ea 
Again, if the potential is simple enough, the above expression may be resummed. See Sec.~\ref{Sec:examples} for an example. Note that this method also applies to analytic potentials, although it might create unnecessary complications.

\subsection{ Effective field theory at leading order }
\label{Sec:EFT-LO}
After obtaining the leading order effective potential according to the prescription of Sec.~\ref{Sec-Veff} one can express the EFT  by an equation of motion for $\psi_s$. Before making the resulting theory explicit, we note that the following important relation exists:
\ba 
\langle V_{\text{int},\psi^*} \rangle  =\langle V_{\text{int}} \rangle_{,\psi_s^*} +\calO(\epsilon^2)= {\mathcal{V}}_{\text{int},\psi_s^*}+\calO(\epsilon^2)\, ,
\label{V'-commute}
\ea 
where $\calO (\epsilon^2)$ indicates that the second (and higher) order corrections in the small operators defined in Eq.~\eqref{epsilons} are neglected. The proof of the first equality can be found in Appendix \ref{App. B} and the second equality follows from the definition Eq.~\eqref{calV_leading}. This relation has two important consequences. First, it allows us to use the effective, coarse-grained potential derived in Sec.~\ref{Sec-Veff} to express the effective equation for $\psi_s$, Eq.~\eqref{psis-eom}, to leading order as follows
\begin{empheq}[box=\tcbhighmath]{equation}
	i \dot{\psi_s} +\dfrac{1}{2m}\laplacian{\psi_s} - \mathcal{V}_{\text{int}}' \psi_s =0+ \calO (\epsilon^2)\, , 
	\label{Schd Eq.}
\end{empheq}
where  $\mathcal{V}_{\text{int}}' \equiv \pdv{\mathcal{V}_{\text{int}}}{\abs{\psi_s}^2}$. The second consequence of  Eq.~\eqref{V'-commute} is that it also allows us to write an effective Lagrangian from which the above effective equation can be derived. Explicitly, we have 
\begin{empheq}[box=\tcbhighmath]{equation}
	\mathcal{L}_{\psi_s,\psi_s^*} = \dfrac{i}{2}(\psi_s^*\dot{\psi_s}-\psi\dot{\psi_s^*})- \dfrac{1}{2m} \grad\psi_s .\grad\psi_s^* 
	- \mathcal{V}_{\text{int}}(|\psi_s|^2) + \calO (\epsilon^2)\, .
	\label{Lag-psis}
\end{empheq}
Comparing this Lagrangian with Eq.~\eqref{Lag-psi} together with Eq.~\eqref{V'-commute} implies that the coarse-graining procedure could have been done at the level of Lagrangian instead of equations of motion. However, note that this works at leading order in the EFT and may be violated at higher orders (see Ref.~\cite{Salehian:2020bon} for an explicit example). Note that the Lagrangian has a global U(1) symmetry; the conserved charge is given by $N\equiv \int |\psi_s|^2 \, \dd[3]{x}$ which is the number of particles. This is expected since, unlike the relativistic theory, particle creation and annihilation do not occur in a non-relativistic theory. As argued in Ref.~\cite{Namjoo:2017nia}, this is a consequence of the fact that each mode in the mode expansion according to Eq.~\eqref{mode exp} carries a distinct and conserved charge. 

The existence of the Lagrangian Eq.~\eqref{Lag-psis} assures that it is possible to study the properties of stationary objects obeying the dynamics governed by the EFT by finding the extrema of the energy of the system. See Sec.~\ref{SecIV}, where we utilize this finding to study solitonic solutions. Eqs.~\eqref{Schd Eq.} and \eqref{Lag-psis} are part of our main results in this paper. 

It is worth noting that the method of coarse-graining is a straightforward tool that enables us to obtain the EFT at leading order in this complex situation. Other methods, such as those discussed in Refs.~\cite{Braaten:2018lmj,Mukaida:2016hwd}, while practical for power-law potentials, are likely to be intractable for more complicated situations that are the main focus of this paper.

\section{Effective fluid description}
\label{SecIII}
In this section, we will obtain an effective fluid theory as an alternative way of describing the non-relativistic scalar field which is particularly useful for cosmology. 
Given $\mathcal{L}_{\phi}$, according to Eq.~\eqref{Lphi},  the energy-momentum tensor reads 
\ba 
T^{\mu}_{\nu} \equiv \pdv{\mathcal{L}_{\phi}}{(\partial_{\mu} \phi)} \partial_{\nu} \phi - \delta^{\mu}_{\nu} \mathcal{L}_{\phi} =-\partial^\mu \phi \, \partial_\nu \phi +\dfrac{1}{2} \delta^{\mu}_{\nu} \left(\partial_\alpha \phi \, \partial^\alpha \phi +  m^2\phi^2 + 2V_{\text{int}} \right)   \, .
\ea 
In analogy with a perfect fluid, one can define the energy density and pressure as follows
\begin{align}
	\label{rho-p-phi}
	 \rho = \dfrac{1}{2} \dot{\phi}^2+ \dfrac{1}{2}(\grad{\phi})^2 + \dfrac{1}{2} m^2 \phi^2 + V_{\text{int}} \, , \qquad 
	 p = \dfrac{1}{2} \dot{\phi}^2 - \dfrac{1}{2}(\grad{\phi})^2- \dfrac{1}{2} m^2 \phi^2 - V_{\text{int}} \, .
\end{align}
From the transformation Eq.~\eqref{full-redef} and using the mode decomposition Eq.~\eqref{mode exp}, it becomes evident that
\begin{eqnarray}
	\label{rho1}
	 \expval*{\rho} = m \abs{\psi_s}^2  +  {\mathcal{V}}_{\text{int}}  +  \dfrac{1}{2m}  \grad{\psi_s} \vdot \grad{\psi^*_s} + \mathcal{O}(\epsilon^2) \ , \label{Av-dens} \\
	 \expval*{p} =  -m (\psi_s \psi_2 + \psi_s^* \psi_{-2}^* )  - {\mathcal{V}}_{\text{int}} -  \dfrac{1}{2m} \grad{\psi_s} \vdot \grad{\psi^*_s}  + \mathcal{O}(\epsilon^2) \label{Av-pres} \, .
	 \hspace{-2.1cm}
		\label{p1}
\end{eqnarray}
As we will soon discover, $\psi_2 \sim \mathcal{O}(\epsilon)$, so all terms in the effective pressure contribute at the working order. The first two terms could have been missed in a naive analysis, assuming $\psi$ is solely composed of the slow-mode. To obtain an expression for $\psi_2 $ we must use Eq.~\eqref{psinu-eom} for $\nu=2$ for which we need   $(V_{\text{int},\psi^*  }  )_{2 }$ to leading order. Similar to Eq.~\eqref{V'-commute}, one can show that 
\ba 
\label{V2-commute}
(V_{\text{int},\psi^*  }  )_{2 } = \psi_s^* \mathcal{V}'_{\text{int}}+\mathcal{O}(\epsilon^2)\, ,
\ea 
 see  Appendix~\ref{App. B} for the proof of this relation. Then, Eq.~\eqref{psinu-eom} yields
\begin{align}
	\label{psi2}
	\psi_2 =  \dfrac{1}{4m^2}\laplacian \psi^*_s -\dfrac{\psi_s^*}{2m}  \mathcal{V}'_{\text{int}} + \mathcal{O}(\epsilon^2) \, .
\end{align}
Recall also that  $\psi^*_{-2} = (\psi_{2})^*$.  Substituting the solution Eq.~\eqref{psi2} into Eq.~\eqref{p1} results in
\begin{align}
	\label{rhoeff}
	& \expval{\rho} =  m \abs{\psi_s}^2  +  {\mathcal{V}}_{\text{int}}  +\dfrac{1}{2m}  \grad{\psi_s}\vdot \grad{\psi^*_s}  + \mathcal{O}(\epsilon^2) \, ,
	\\
	 \label{peff}
&	 \expval{p} = \abs{\psi_s}^2   {\mathcal{V}}_{\text{int}}'  - {\mathcal{V}}_{\text{int}} - \dfrac{\laplacian}{4m} |\psi_s|^2 \, .
\end{align}

Having found the coarse-grained energy density and pressure, for  cosmological applications, we express the results for a homogeneous and isotropic background and linear fluctuations. That is, define $ \psi(t,\vb{x}) = \bar{\psi}(t) + \delta\psi (t,\vb{x})$ which results in $\rho(t,{\bf x})=\bar \rho(t)+\delta \rho(t,{\bf x})$ and $p(t,{\bf x})=\bar p(t)+\delta p(t,{\bf x})$   and assume $|\bar \psi | \gg |\delta \psi|$.  Then, from Eqs.~\eqref{rhoeff} and \eqref{peff} for the background (homogeneous and isotropic) quantities we have  
\ba 
\expval{\bar \rho} =  m \abs{\bar \psi_s}^2  + \bar{\mathcal{V}}_{\text{int}}  + \mathcal{O}(\epsilon^2) 
\, , \qquad 
\expval{\bar p} =\abs{\bar \psi_s}^2  \, \bar{\mathcal{V}}_{\text{int}}'  -\bar{\mathcal{V}}_{\text{int}}  + \mathcal{O}(\epsilon^2) \, ,
\label{rho_p_background}
\ea 
whereas, for the linear perturbations, we have 
\begin{align}
	& \expval{\delta\rho} =  (m +\bar{\mathcal{V}}_{\text{int}}')( \bar{\psi}_s  \delta \psi_s^* + \bar{\psi}^*_s \delta \psi_s ) + \mathcal{O}(\epsilon^2)  \, , \\
	& \expval{\delta p} = \big( \abs{\bar{\psi}_s}^2 \bar{\mathcal{V}}_{\text{int}}''  - \dfrac{\nabla^2}{4m}\big)( \bar{\psi}_s  \delta \psi_s^* + \bar{\psi}^*_s \delta \psi_s ) + \mathcal{O}(\epsilon^2) \, . 
\end{align}
In the above equations, we defined $\bar{\mathcal{V}}_{\text{int}}\equiv {\mathcal{V}}_{\text{int}}(|\bar \psi_s|^2)$; likewise for derivatives. An important quantity that can be obtained from these results is the sound speed, i.e., the speed of propagation  of small fluctuations in the medium. We have 
\begin{equation}
	\label{cs2-flat}
	c_s^2 \equiv  \dfrac{\delta p}{\delta\rho} = \dfrac{k^2}{4 m^2} + \abs{\bar{\psi}_s}^2 \dfrac{\bar{\mathcal{V}}_{\text{int}}''}{m} + \mathcal{O}(\epsilon^2) \, ,
\end{equation}
where we presented the result in the Fourier space which is appropriate for studying linear fluctuations. Note that to obtain both terms in the sound speed it was necessary to take into account the contribution from $\psi_2$. The first term is well-known to arise as a result of the quantum pressure (see, e.g., Ref.~\cite{Hwang:2009js}) and the second term generalizes the effect of a quartic self-interaction (derived, e.g., in Ref.~\cite{Salehian:2020bon}) to an arbitrary potential.

As a validity check of the obtained sound speed, one can derive a second order equation for fluctuations, where the sound speed clearly appears. First, note that the equations for the background field and linear fluctuations are given by
\begin{align}
		& \, \, i \dot{\bar{\psi_s}} =  \bar{\mathcal{V}}_{\text{int}}'  \bar{\psi_s}  \, , 
		\label{BPsi Eq.} \\
		& i \delta \dot{\psi_s}  = - \dfrac{1}{2m} \nabla^2   \delta\psi_s + \bar{\mathcal{V}}'_{\text{int}} \delta\psi_s   + \bar{\mathcal{V}}_{\text{int}}''  \left( \bar{\psi_s}\, \delta\psi_s^* + \bar{\psi_s}^*\delta\psi_s \right) \bar{\psi_s} \, ,
		\label{PPsi Eq.}
\end{align}
where, from now on, we omit the $\mathcal{O}(\epsilon^2)$ symbol since we always work up to this order. 
Next, define the density contrast by $\delta \equiv  \expval{\delta\rho}/\expval{\bar{\rho}}$. Up to the working order, we have 
\ba 
\delta = \dfrac{1}{m|\bar \psi_s|^2}(m +\bar{\mathcal{V}}_{\text{int}}')( \bar{\psi}_s  \delta \psi_s^* + \bar{\psi}^*_s \delta \psi_s ) \, .
\ea 
Using Eqs.~\eqref{BPsi Eq.} and \eqref{PPsi Eq.} and their time derivatives one can obtain a second order equation of motion for $\delta$ in Fourier space as follows
\begin{equation}
\ddot{\delta} + c_{s}^2 \, k^2  \, \delta = 0 \, ,
\end{equation}
where $c_s^2$ here matches exactly with Eq.~\eqref{cs2-flat}. While we obtained consistent sound speed at this order from both considerations, it is worth stressing that this may not hold at higher orders, especially when gravitational effects are taken into account. If this happens, it is evident that the equation of motion gives the correct sound speed. This is because the relation $c_s^2 \equiv   \dfrac{\delta p}{\delta\rho}$ relies on an analogy between the scalar field and the perfect fluid theories. Generally, such an analogy may not be exact at higher orders in the EFT. In fact, in Ref.~\cite{Salehian:2020bon} it is shown that in the linear cosmological perturbation theory (and at higher order in the EFT), not only the correct sound speed deviates from the perfect fluid counterpart, $c_s^2=\delta p/\delta \rho$, but also an accurate fluid description requires introducing other quantities such as viscosity which do not appear in the theory of a perfect fluid. This primarily arises from the coupling of the field to gravity that induces additional backreaction effects.

\section{Effective equations in the expanding background}
\label{Sec:EFT_FRW}
 
Extending our results to an expanding background is straightforward as long as the leading order effective theory is concerned. For this purpose, we combine the results we obtained so far with the results of Ref.~\cite{Salehian:2020bon} where the effective equations in the expanding background (but only with a quartic self-interaction) are presented. Consider an expanding homogeneous and isotropic universe and linear perturbations around it. In Newtonian gauge, the metric takes the following form \cite{Weinberg:2008}
\begin{equation}
\dd{s}^2 = -(1+2\Phi)\dd{t}^2 + a^2 (1-2\Phi) \delta_{ij}\dd{x^i}\dd{x^j}\, ,
\end{equation}
where $a=a(t)$ is the scale factor and $\Phi=\Phi(t,{\bf x})$ is the gravitational potential that quantifies linear fluctuations of the metric in this gauge.  Note that the equality of the fluctuations of the metric's temporal and spatial components is the straight result of vanishing anisotropic stress tensor. This restriction is valid for a dark matter-only universe but needs to be revised to, for example, account for the contribution of neutrinos.

For the background variables and for the linear perturbations in Newtonian gauge we have 

\begin{empheq}[box=\tcbhighmath]{gather}
  i \dot{\bar{\psi_s}} +\dfrac{3}{2}i H \bar \psi_s -  \bar{\mathcal{V}}_{\text{int}}'  \bar{\psi_s}  =0\, , \qquad 3 \Mpl^2 H^2=m|\bar \psi_s|^2+ \bar{\mathcal{V}}_{\text{int}} \, ,
  \label{Schrodinger-FRW}
\\
 i \dot{\delta {\psi_s}} +\dfrac32 i H \delta \psi_s + \dfrac{\nabla^2 \delta\psi_s }{2m a^2}    -m\,  \bar \psi_s \Phi - \bar{\mathcal{V}}'_{\text{int}} \delta\psi_s   - \bar{\mathcal{V}}_{\text{int}}'' \bar{\psi_s}  \left( \bar{\psi_s}\delta\psi_s^* + \bar{\psi_s}^*\delta\psi_s \right) =0\, ,
\\
 \dfrac{\nabla^2 \Phi}{a^2} = \dfrac{m}{2\Mpl^2} \left(\bar{\psi_s}\delta\psi_s^* + \bar{\psi_s}^*\delta\psi_s   \right)\, ,
\end{empheq}
where $H=\dot a/a$ is the Hubble parameter and $\Mpl \equiv 1/\sqrt{8\pi G}$ is the reduced Planck mass. The Schr\"odinger-like equations are derivable from applying the field redefinition Eq.~\eqref{full-redef} to the  Klein-Gordon equation in an expanding universe and with a generic potential. The Friedmann equation could be guessed as on the right-hand side the background energy density must appear. Likewise, the Poisson equation has the linear density perturbation on the right-hand side, as it should.

In the fluid language, the background quantities are given by
\begin{empheq}[box=\tcbhighmath]{equation}
\expval{\bar \rho} =  m \abs{\bar \psi_s}^2  + \bar{\mathcal{V}}_{\text{int}}   
\, , \qquad 
\expval{\bar p} =\abs{\bar \psi_s}^2  \, \bar{\mathcal{V}}_{\text{int}}'  -\bar{\mathcal{V}}_{\text{int}}  \, .
\label{rho-p-FRW}
\end{empheq}
For the linear perturbations, we present the results for the comoving density contrast in Fourier space. It is well-known that for a perfect fluid, this quantity obeys \cite{Salehian:2020bon}
\begin{empheq}[box=\tcbhighmath]{equation}
	\label{delta}
\ddot \delta +2 \gamma H \dot \delta + c_s^2 \dfrac{k^2}{a^2} \delta = \dfrac32 H^2 \kappa \, \delta\, ,
\end{empheq}
where $c_s^2$ is the sound speed and
\ba 
\gamma \equiv  1-\dfrac{3p}{\rho}+\dfrac{3\dot p}{2\dot \rho}\, , \qquad \kappa \equiv 1+\dfrac{8p}{\rho}-\dfrac{6\dot p}{\dot \rho} \, .
\ea 
Using Eqs.~\eqref{cs2-flat} and \eqref{rho-p-FRW} it is easy to obtain the following relations in an expanding background 
\begin{empheq}[box=\tcbhighmath]{gather}
	 \label{cs2}
	c_{s}^2 =  \dfrac{k^2}{4 m^2 a^2}  + \dfrac{1}{m} \abs{\bar{\psi}_s}^2   \bar{\mathcal{V}}_{\text{int}}'' \, ,
	\\
	\label{gamma}
\gamma = 1+\dfrac{3}{2m|\bar \psi_s|^2}  \left( 2 \bar{\mathcal{V}}_{\text{int}} -2 |\bar \psi_s|^2 \bar{\mathcal{V}}_{\text{int}}'+|\bar \psi_s|^4\bar{\mathcal{V}}_{\text{int}}'' \right)\, ,
\\
\label{kappa}
\kappa = 1-\dfrac{2}{m|\bar \psi_s|^2}  \left( 4 \bar{\mathcal{V}}_{\text{int}} -4 |\bar \psi_s|^2
 \bar{\mathcal{V}}_{\text{int}}'+3|\bar \psi_s|^4\bar{\mathcal{V}}_{\text{int}}'' \right) \, .
\end{empheq}
These are our other key findings in this paper. 

In a pure matter-dominated universe, we have $\bar p=0$, $\bar \rho \sim a^{-3}$ and $c_s=0$ which imply a linear growth of the density contrast, $\delta \sim a$. If dark matter can be described by a non-relativistic field, due to the non-zero pressure and sound speed, the evolution would slightly differ. In particular, the self-interaction may induce an effective sound speed that becomes imaginary in a range of scales, leading to an exponential growth of structures. Other sources of deviations from pure matter domination exist due to deviations of $\gamma$ and $\kappa$ from unity.

\section{Examples}
\label{Sec:examples}
In this section, we examine our method for deriving EFT through a few explicit examples.  As mentioned above, although the proposed method applies to a general potential, we focus on these examples because of their relevance to cosmological applications and the possibility of comparing our results with earlier work.
To simplify notation, we define $\psi_0 \equiv |\phi_0|\sqrt{m/2}$, where $\phi_0$ is a quantity used in different potentials (and thus may have different meanings, but always has a mass dimension of 1). Additionally, we define $x \equiv |\psi_s|/\psi_0$ to analyze the asymptotic behavior of the coarse-grained potentials. 

\subsection*{$\bullet \, V_{\text{int}}=\dfrac{\lambda}{4!}\phi^4$ }
This  is a well-known but still illuminating example.
For this potential, we have\footnote{One may consider a more general power-law potential of the form
	$  V_{\text{int}}=V_0 \ \big(\frac{\phi}{\phi_0} \big)^{2\beta}$ for an arbitrary $\beta$. In this case, one obtains  $ {\mathcal{V}}_{\text{int}} =\mathcal{V}_0 \  \big(\frac{\abs{\psi_s}}{\psi_0} \big)^{2\beta}  $ where
	$ \mathcal{V}_0 \equiv V_0 \ \pi^{-1/2} \ \Gamma(\frac{1}{2} + \beta) \Gamma(1 + \beta)^{-1}$.} 
\ba 
{\mathcal{V}}_{\text{int}} = \dfrac{\lambda}{16 m^2} \abs{\psi_s}^4\, ,
\ea  
from which we obtain
\begin{eqnarray}
	\expval{\bar \rho}=m|\bar \psi_s|^2+\dfrac{\lambda}{16 m^2} \abs{\bar \psi_s}^4\, , \qquad  
	\expval{\bar p}=\dfrac{\lambda}{16 m^2} \abs{\bar \psi_s}^4\, ,\quad 
	\\
	c_s^2 =  \dfrac{k^2}{4 m^2a^2} + \dfrac{\lambda \abs{\bar{\psi}_s}^2}{8 m^3}\, , \qquad 
	\gamma =1 \, , \qquad \kappa = 1- \dfrac{\lambda \abs{\bar{\psi}_s}^2}{4 m^3}\, .
\end{eqnarray}
These results are consistent with the findings of Ref.~\cite{Salehian:2020bon}. Note that if one naively neglects the contribution of $\psi_2$ to the effective pressure, the last term of Eq.~\eqref{rho-p-phi} might suggest  $\expval{\bar p}\to -\frac{\lambda}{16 m^2} \abs{\psi_s}^4$, which carries the wrong sign. This could have significant consequences, e.g., when one aims to modify the expansion history of the universe in a certain way (see Ref.~\cite{Moss:2024mkc} for an example).

\subsection*{$\bullet \, V_{\text{int}}=V_0 \, \big(\frac{\phi}{\phi_0}\big)^4 \ln \left(\frac{\phi^2}{\phi_0^2}\right)$ }
This non-analytic potential may arise due to radiative corrections similar to the Coleman-Weinberg model. For this model we have
\ba 
\label{Veff_CW}
\calV =  {\cal V}_0\, \big(\dfrac{|\psi_s|}{\psi_0}  \big)^{4}\,  \ln  \big(\varrho_0 \dfrac{|\psi_s|^2}{\psi_0^2}  \big)\, ,
\ea 
where 
\ba 
{\cal V}_0=\dfrac{3}{8} V_0\, , \qquad 
 \varrho_0 = \dfrac{1}{4} \exp(\frac{7}{6}) \, .
\ea 

\begin{figure}
		\includegraphics[scale=.73]{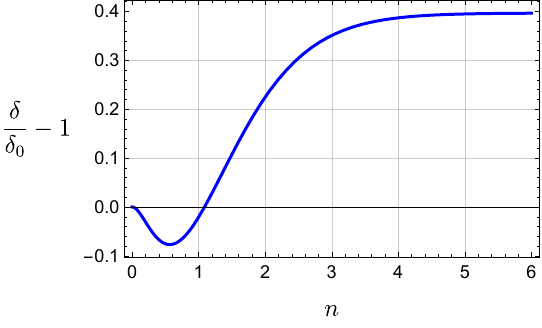}
		\hspace{1cm}
	\includegraphics[scale=.65]{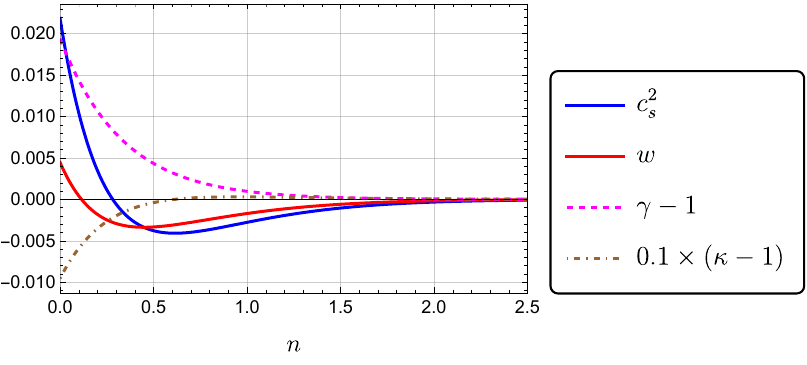}
	\caption{The behavior of the density contrast (compared to the pure matter-domination predictions) and other physical quantities for the potential Eq.~\eqref{Veff_CW}. The horizontal axes represent the number of efolds defined by $n=\ln \big( a(t)/a(0)\big)$. For this plot we have set $m=10^{-5}\, $eV, ${\cal V}_0 =2 \times 10^{-7}\, $eV$^4$, $\psi_0=1\,$eV$^{3/2}$, $k=1/$Mpc, and for the initial conditions at $n=0$, we have set  $\bar \rho(0) = 1.5 \times 10^{-24}\, g/{\text{cm}}^3$, $\delta(0)=\delta_0(0)=1$ and $\delta'(0)=1$. Note that we multiplied $\kappa-1$ by a factor of $0.1$ to make all curves visible in a single graph.}
	\label{fig:delta}
\end{figure}

An interesting feature of this model, particularly when applied to cosmology, is that the potential can switch the sign, leading to an instability for a certain range of time. For this model, we find it constructive to perform a numerical analysis of the behavior of physical quantities in an expanding background. Since deviations from a pure matter-dominated universe cannot be significant, we assume that the background energy density approximately behaves like $\bar \rho \propto a^{-3}$, similar to the pure matter energy density. Then, we replace $|\bar \psi_s|^2$ in Eq.~\eqref{rho-p-FRW} and Eqs.~\eqref{cs2}-\eqref{kappa} with $\bar \rho(t)/m$ to obtain the behavior of the equation of state and other parameters that appear in Eq.~\eqref{delta}, the equation of motion for the density contrast, which we solve numerically. See Fig.~\ref{fig:delta} for the behavior of different quantities for a (not necessarily realistic) choice of parameters and initial conditions. In the left panel of Fig.~\ref{fig:delta} we plot the fractional difference between $\delta$ from this theory and the pure matter-domination predictions. We denote the density contrast in a pure matter-dominated universe by $\delta_0$, which is well-known to behave like $\delta_0 \sim a$. The effect of the sign flip in different parameters is vivid in the figures.  

\subsection*{$\bullet \, V_{\text{int}}=V_0 \cos(\frac{\phi}{\phi_0}+{\scriptstyle  \theta_0})$ }
In this example, $V_0$, $\phi_0$, and $\theta_0$ are constant parameters.\footnote{As we were working on this paper, Ref.~\cite{GalazoGarcia:2024fzq}, appeared that studies a specific potential that resembles the potential we considered in this example. We agree with the coarse-grained potential obtained in Ref.~\cite{GalazoGarcia:2024fzq}.}  This form of potential may be generated from non-perturbative effects for axion-like particles.\footnote{This form of potential is typically regarded as also responsible for the mass term, when it is expanded in the small field limit. Here, we assume a distinct and dominant mass term that enables us to investigate this potential in the large field (but still non-relativistic) limit. We take a phenomenological approach and do not attempt to provide theoretical justification for this assumption.}
In this case, from the resummation of Eq.~\eqref{V_expand_analytic} we obtain 
\ba 
{\mathcal{V}}_{\text{int}} =V_0 \cos(\theta_0) \, J_0(\dfrac{|\psi_s|}{\psi_0})\, , 
\ea 
where $J_0(\cdot)$ is the Bessel function of the first kind. To gain some intuition from this result, we may expand the coarse-grained potential in the large field limit, which yields ${\mathcal{V}}_{\text{int}} \propto  \cos(x-\frac{\pi}{4})/\sqrt{x}$. We observe that, while the original potential oscillates in the range $[-V_0,V_0]$, the amplitude of the oscillations of the coarse-grained potential decreases as the amplitude of the non-relativistic field increases. This occurs  because the coarse-graining operation diminishes the potential more effectively when the frequency of oscillations $\dot \phi/\phi_0 \sim m \phi/\phi_0$ increases. As a final remark, note that the potential $V_{\text{int}}=V_0 \sin(\frac{\phi}{\phi_0})$ --- which can be obtained by choosing $\theta_0 = -\pi/2$ --- vanishes after coarse-graining, to leading order in the  NREFT (since only odd powers of $\phi$ appear in the Taylor expansion of the potential). 

\subsection*{$\bullet \, V_{\text{int}}=V_0 \, e^{\pm \frac{\phi}{\phi_0}}$ }
For this dilaton-like potential, in the non-relativistic limit, it makes no significant difference whether a positive or negative sign is chosen in the exponent. In fact, irrespective of the sign, we obtain 
\ba 
{\mathcal{V}}_{\text{int}} =V_0 \,  I_0(\dfrac{|\psi_s|}{\psi_0})\, ,
\ea  
where $I_0(\cdot)$ is the modified Bessel function of the first kind.\footnote{The difference arising from the choice of the sign in the exponent is expected to appear at higher orders in the EFT. It is also worth noting that if the power of the field in the exponent is even, the sign does matter, even at the leading order EFT. For example, consider a potential of the form  $ V_{\text{int}}= V_0 \ \exp ({\pm \frac{\phi^2}{\phi_0^2}} )$ which yields  $ {\mathcal{V}}_{\text{int}} = V_0 \ \exp(\pm \frac{\abs{\psi_s}^2}{2\psi_0^2}) I_0(\dfrac{\abs{\psi_s}^2}{2\psi_0^2} )$. In this case, the asymptotic behavior is like $e^{2x}/\sqrt{x}$ and $1/\sqrt{x}$ for the positive and negative signs, respectively. Interestingly, we observe that for the negative sign, the coarse-grained potential is suppressed only by a power-law (rather than exponentially) in the large field limit.} The asymptotic behavior of this potential is given by ${\mathcal{V}}_{\text{int}} \propto e^x/\sqrt{x}$. The exponential factor is expected but we again see the additional suppression factor like $1/\sqrt{x}$ due to coarse-graining.

\section{Solitonic solutions}
\label{SecIV}
In this section, we present the equations necessary for studying non-relativistic solitonic solutions.\footnote{In this paper, we refer to any stationary solution that is formed by balancing the self-interaction and other forces as ``soliton”. Depending on which forces are at work, other names are given in the literature such as boson star or axiton.} We neglect the effect of expansion, work in the weak gravity limit, and consider the leading order effective theory. For relativistic corrections to our leading order effective theory, see Ref.~\cite{Salehian:2021khb}. Generalising Eq.~\eqref{Schd Eq.} to include the effect of gravity yields the Schr\"odinger--Poisson system of equations:
\ba 
i \dot{\psi_s} +\dfrac{1}{2m}\laplacian{\psi_s} -m \, \Phi_{_N}\, \psi_s - \mathcal{V}_{\text{int}}' \psi_s =0 \, 
, \qquad
\nabla^2 \Phi_{_N} = 4 \pi G \, m |\psi_s|^2 \, ,
	\label{Schd-GR}
\ea 
where $\Phi_{_N}$ is the Newtonian gravitational potential. The above coupled system of equations may be solved numerically to study the soliton formation. However, since solving partial differential equations is typically intensive, we will also present two simplified methods for examining the properties of solitonic solutions in the remainder of this section.

\subsection{Soliton's profile}
\label{Sec:profile}

To search for a stationary solution, one may use an ansatz of the form $\psi_s = f(r, \theta , \phi) \  e^{-i \mu t}$ --- where $\mu$ is a constant and $f= f(r, \theta , \phi)$ is a real function --- to  simplify the set of equations as follows: 
\ba 
\label{eq_for_profile}
\mu f +\dfrac{1}{2m}\laplacian{f} -m \, \Phi_{_N}\, f - \mathcal{V}_{\text{int}}' f =0 \, 
, \qquad
\nabla^2 \Phi_{_N} = 4 \pi G \, m f^2 \, ,
\ea 
where $\mathcal{V}_{\text{int}}$ must be understood as a function of $f^2$. The price of this simplification is loosing  the dynamics of the system. However, it allows one to determine the behavior of the soliton's profile. 

It has been argued that in many cases, an exponential ansatz suffices to approximate the profile despite the fact that it cannot be correct around the center of the soliton (since we need $f \to 0$ for $r \to 0$ to avoid singularity). See, e.g., Ref.~\cite{Schiappacasse:2018} for a discussion on this matter. However, this statement is not necessarily true for more complex, non-power-law potentials, which are the main focus of this paper. We study a few examples to demonstrate that strong deviation from exponential can occur in this situation. 

We further simplify the problem by neglecting gravity and assuming spherical symmetry. Thus, our goal is to solve the following ordinary differential equation for $f=f(r)$:
\begin{equation}
	\label{eq1}
	\dv[2]{f(r)}{r}
+ \dfrac{2}{r} \dv{f(r)}{r} - \mathcal{V}_{\text{int}}' f(r) + 2 m \mu f(r) = 0 \, .
\end{equation}
We solve this equation for a number of potentials. To find the solution numerically, we iteratively refined the value of the parameter $\mu$ by requiring that the resulting solution $f(r)$ is a smooth and localized function. The initial conditions are  $f(0)=f_0$, for a choice of $f_0$, and $f'(0)=0$, to ensure the singularity at the center is avoided. 

\begin{figure}
	\centering
	
	\includegraphics[width=6.5cm]{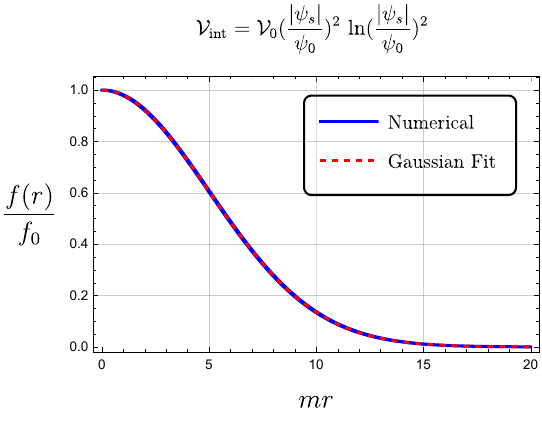}

	\caption{Soliton's profile for a quadratic potential, altered by a logarithmic factor. The blue curve shows the  numerical result and the dashed red curve shows the best Gaussian fit. For this plot we have set $m$, $\psi_0$, and $|\mathcal{V}_0| $ the same as those in Fig.~\ref{fig:delta} but have chosen  $\mathcal{V}_0 <0$ to have a localized solution. We have also set $f_0 = 100 \ \rm{eV^{3/2}} $.}
	\label{fig:psi2}
\end{figure}

As a simple comparison between power-law and non-power-law potentials, first consider $ \mathcal{V}_{\text{int}} = \mathcal{V}_0 (\frac{ |\psi_s|}{\psi_0})^2 $. This is like a mass term and can indeed be removed by rescaling the mass. However, we study it because its simplicity allows us to clearly observe how the profile is affected by a modification of the potential. For this potential, the system admits analytical solutions of the exponential form $ \exp(\pm \frac{r}{R_0})$, where $R_0= \frac{\psi_0}{\sqrt{2m( \mu \psi_0^2 - \mathcal{V}_0)}}$. Now consider a simple modification of the potential by adding a logarithmic factor, i.e., $ \mathcal{V}_{\text{int}} = \mathcal{V}_0 (\frac{ |\psi_s|}{\psi_0})^2 \ln (\frac{|\psi_s|^2}{\psi_0^2})$. In Fig.~\ref{fig:psi2}, we show the behavior of the profile for this potential. Interestingly, after adding the logarithmic factor, the corresponding soliton’s profile transitions to a Gaussian-like behavior, indicating a significant change in the structure of the soliton.

\begin{figure}
	\centering
	\includegraphics[width=6.5cm]{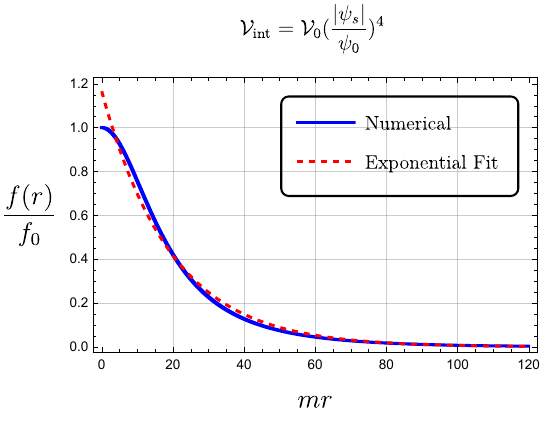}
	\hspace{1cm}
	\includegraphics[width=6.5cm]{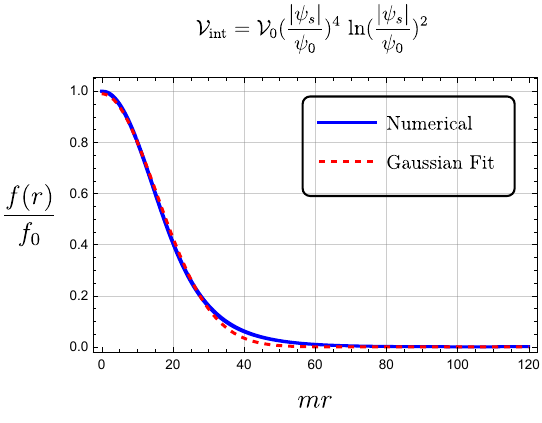}

	\caption{Soliton's profile for $\mathcal{V}_0 (\frac{|\psi_s|}{\psi_0})^4 $ with and without the logarithmic factor.  For these plots we have set $m$, $\psi_0$ and $|\mathcal{V}_0| $  the same as those in Fig.~\ref{fig:delta} and $f_0 = 0.5 \ \rm{eV^{3/2}} $. For the left panel we have considered $\mathcal{V}_0 <0$ to have a localized solution while  for right panel  we have $ \mathcal{V}_0 >0$. (The sign difference is compensated by the logarithmic factor.) }
	\label{fig:psi4}
\end{figure}

As another example, in Fig.~\ref{fig:psi4}, we compare the profile of the soliton for the potentials of the form $ \mathcal{V}_{\text{int}} = \mathcal{V}_0 (\frac{|\psi_s|}{\psi_0})^4 $ and $ \mathcal{V}_{\text{int}} = \mathcal{V}_0 (\frac{ |\psi_s|}{\psi_0})^4 \ln (\frac{|\psi_s|^2}{\psi_0^2})$. We again see that the profile tends to become Gaussian after the inclusion of the logarithmic factor.\footnote{See also Ref.~\cite{GalazoGarcia:2024fzq}, which considers a different non-polynomial potential, for which a Gaussian profile is again found. }


\subsection{Energy balance analysis}

The second approximate analysis permitted by our EFT framework is to examine the balance among various forces at play in the solitonic solution. This analysis, while still significantly simpler than solving Eq.~\eqref{Schd-GR}, enables the investigation of the stability of the stationary solutions discussed in Sec.~\ref{Sec:profile} and allows for the exploration of the mass-radius relation for solitons. 

 We have shown in Sec.~\ref{Sec:EFT-LO} that the leading order NREFT can be expressed in terms of a Lagrangian as given in Eq.~\eqref{Lag-psis}. This ensures that a consistent Hamiltonian exists at this order, from which one can analyze the existence and stability conditions of the stationary solutions. Taking into account gravity, the total energy of the system in the stationary phase consists of four terms as follows 
 \begin{equation}
 	\label{Hamiltonian}
 	H_{\text{T}} = H_{\text{m}}  + H_{\text{grad}} + H_{\text{int}}  + H_{\text{grav}} \, ,
 \end{equation}
 where $H_{\text{m}},  H_{\text{grad}} , H_{\text{int}}$, and $H_{\text{grav}}$ are the rest mass energy, gradient energy,  self-interaction energy, and garvitational energy,  respectively, and are given by
\[	 H_{\text{m}} \equiv m \int \dd[3]{r} \abs{\psi_s}^2 \ , \quad 
 	 H_{\text{grad}} \equiv \dfrac{1}{2m} \int \dd[3]{r} \grad{\psi_s} \vdot \grad{\psi_s^*} \, , \quad 
 	 H_{\text{int}} \equiv \int \dd[3]{r} \mathcal{V}_{\text{int}}  \, , \]
 \begin{equation}
 		 \label{energies}
 	 H_{\text{grav}}  \equiv  -\dfrac{Gm^2}{2} \int \dd[3]{r} \int \dd[3]{r'} \dfrac{\abs{\psi_s({\bf r})}^2 \abs{\psi_s({\bf r'})}^2}{|{\bf r}-{\bf r'}|} \, .
 \end{equation}

To proceed, one needs to make an assumption about the soliton's profile. Motivated by the results of Sec.~\ref{Sec:profile}, we consider a generic profile as follows:
\ba 
\label{profile}
\psi_s(r) =c_0 \sqrt{ \dfrac{N}{ R^3}}\, e^{-(\frac{r}{R})^q}\, ,
\ea 
where $N$ is the number of particles, $R$ is the size of the soliton\footnote{More precisely, $R$ is the radius at which the field amplitude  is an efold less than the peak.}, $q$ is an arbitrary positive constant and 
\ba 
c_0^2 = \dfrac{q \, 2^{\frac{3}{q}-2}}{\pi \Gamma(\frac{3}{q})}\, .
\ea   
The normalization factor is determined by the constraint $\int |\psi_s|^2 \, \dd[3]{r} = N$. The value of $q$  depends on the choice of potential. However, we will see that the different choices of $q$ only affect the numerical factors, not the overall scalings with physical quantities. With this choice of profile, one can perform the integrations in Eq.~\eqref{energies} to obtain the total energy of the system as follows:
\ba 
\label{energy}
E = N m +c_1 \dfrac{N}{mR^2} -c_2 \dfrac{G m^2 N^2}{ R} + 4 \pi R^3 \, \calW(N/R^3)\, ,
\ea 
where
\ba 
\label{c1-c2}
c_1 = 2^{\frac{2}{q}-3} \, q^2 \, \frac{ \Gamma (2+\frac{1}{q} )}{ \Gamma  (\frac{3}{q} )}
\, , \qquad 
c_2 =\dfrac{2^{\frac{6}{q}} q^2}{\Gamma^2(\frac{3}{q})}\int \dd{\tilde{r}} \int \dd{\tilde{r}'} \, \frac{ \tilde{r}^2 \tilde{r}'^2 \, e^{-2 \left(\tilde{r}^q+\tilde{r}'^q\right)}}{(\tilde{r}+\tilde{r}') + |\tilde{r}-\tilde{r}'| } \, .
\ea 
The integrals in the definition of $c_2$ can be performed, the resulting expression may be found in Appendix~\ref{App.c2}.
In Eq.~\eqref{energy}, we have also defined
\ba 
 \calW(\hat x) = \int_0^{\infty} \calV \big(\hat x\, c_0^2 e^{-2\tilde r^q}  \big)\, \tilde r^2 \dd{\tilde r}\, ,
\ea 
where, recall that $\calV$ is understood as a function of $|\psi_s|^2$ (hence its argument is made explicit using Eq.~\eqref{profile}). 
Note that $\calW$  depends on $R$ and $N$ only through the combination $\hat x=N/R^3$. If the potential is too complex to allow one to directly perform the above integration, one might use the series expansion such as Eqs.~ \eqref{calV_anal} and \eqref{V_expand_analytic}. For example, from Eq.~\eqref{calV_anal} we obtain\footnote{In this relation, we assume $\alpha_0=0$ since this corresponds to  a non-physical divergent term for a soliton.}
\begin{equation}
	\calW(\hat x) =\frac{1}{q }\,\Gamma(\frac3q)   \sum_{n=1}^{\infty}   \alpha_{2n} \,   (2n)^{-\frac3q}\, \binom{2n}{n}\, \left( \frac{\hat x \, c_0^2 }{ {2 m }}\right)^n \, .
\end{equation}  
A compact form for $\calW$ may be obtained if either resummation or truncation is possible.

A localized object can be formed if, for fixed $N$, the energy has an extremum as a function of $R$. That is, we require $\partial E/\partial R=0$ which results in
\ba 
\label{E'}
-m R^3 \,  \dfrac{\partial E}{\partial R}=2 c_1 N- c_2\, G m^3 N^2 R-12 \pi \,  m \, R^5\,  \calW+12 \pi \,  m\,  N\, R^2 \, \calW' =0\, ,
\ea 
where $\calW' =\big(\partial \calW(\hat x)/\partial \hat x \big)|_{\hat x=\frac{N}{R^3}}$.
The existence of a solution to the above equation must be accompanied by the stability condition. We have
\ba 
\label{E''}
\dfrac{1}{2}m R^5 \dfrac{\partial^2 E}{\partial R^2} = 
3 c_1 N R -  c_2 G m^3 N^2 R^2+18 \pi  m N^2 \calW''-12 \pi  m N R^3 \calW'+12 \pi  m R^6 \calW \, ,
\ea 
and we require $\dfrac{\partial^2 E}{\partial R^2}>0$ for the stability of the solution. The above expressions will be particularly useful for studying the mass-radius relation of solitons, which will be explored elsewhere \cite{no-go}.

\section{Conclusion}
\label{Sec:conclusion}

In this work, we developed a non-relativistic effective field theory (NREFT) based on a relativistic, self-interacting scalar field theory with a general potential. In the Minkowski spacetime, the Lagrangian that describes our NREFT is given by Eq.~\eqref{Lag-psis}.  A key feature of this framework is its ability to accommodate a wide range of potentials, including non-power-law potentials or  non-analytic ones around the classical vacuum. The procedure for deriving the effective potential for each class is detailed in Secs.~\ref{Sec_anal} and \ref{Sec_non-anal}, respectively. While NREFTs are usually viewed as appropriate in the limit of low velocities, weak couplings, and small field amplitudes, our approach relaxes the latter assumption, enabling the consideration of large field amplitudes and treating them in a non-perturbative manner.  

From the resulting NREFT, we have demonstrated that a fluid description can be attributed to the scalar field, a procedure of particular interest in cosmology. This approach will be especially useful, for example, in describing ultra-light dark matter. The influence of self-interactions in an expanding universe can be tracked through the physical variables that quantify the properties of a perfect fluid, such as the energy density, pressure, and sound speed — see Eqs.~\eqref{rho-p-FRW}, \eqref{cs2}-\eqref{kappa}. The existence of self-interaction may affect the dynamics of the universe and its matter content, leaving observable imprints through cosmological phenomena such as structure formation, as evident from the equation of motion for the density contrast, Eq.~\eqref{delta}. As shown in Ref.~\cite{Salehian:2020bon}, deviations from a perfect fluid are expected to appear at higher orders in the NREFT, the exploration of which is worthwhile but beyond the scope of this paper.

With the established NREFT framework, we are also equipped to study solitonic solutions. We have presented the equations governing the dynamics and formation of solitons (Eq.~\eqref{Schd-GR}) and the simplified equations suitable for studying the soliton’s profile (Eq.~\eqref{eq_for_profile}), along with the analysis of its existence and stability conditions (Eq.~\eqref{Hamiltonian}). In several concrete examples, we observed that for more complex potentials, the soliton's profile tends to admit a Gaussian fit rather than an exponential. The generality of this observation remains an intriguing open question.

This NREFT framework facilitates the study of various phenomena --- including the core-cusp problem, structure formation, and cosmic microwave background anisotropies --- when dark matter consists of ultra-light particles.  In particular, the framework developed here allows one to prove a no-go theorem for the mass-radius relation of solitons, implying that the core-cusp problem does not have a simple solution in the context of ultra-light dark matter modes. This result will be presented in Ref.~\cite{no-go}. It would be also interesting to apply the same method to establish the NREFT when multiple fields contribute to the dynamics of such phenomena,  which we also leave for future work.

\appendix
\renewcommand{\thesection}{\Alph{section}} 
\renewcommand{\theequation}{\Alph{section}.\arabic{equation}}

\section{Proof of some useful relations} \label{App. B}
\setcounter{equation}{0} 

In this appendix, we briefly outline the proof of two relations we used in the main text. 

\subsection{Eq.~\eqref{V'-commute}: $	\expval{ V_{\text{int},\psi^*} } = {\mathcal{V}_{\text{int}}}_{,\psi_s^*} + \mathcal{O}(\epsilon^2)$}
We start with Eq.~\eqref{V'-commute} which, roughly, states that the coarse-graining and derivative with respect to the field commute, to leading order in the NREFT.

For an analytic potential, the identity
\begin{equation}
	V_{\text{int},\psi^*} = \dfrac{e^{imt}}{\sqrt{2m}} V_{\text{int},\phi} \, ,
\end{equation}
along with the series expansion Eq.~\eqref{expand_analytic} result in
\begin{equation}
	V_{\text{int},\psi^*} = 
	\sum^{\infty}_{n=0} \dfrac{V_{\text{int}}^{(n+1)}}{n!} \dfrac{e^{imt}}{(\sqrt{2m})^{n+1}} \sum^{n}_{k=0} \binom{n}{k} e^{imt(n-2k)}  \psi^k  \psi^{* \ n-k} \, .
\end{equation}
Omitting the oscillatory terms at leading order yields
\begin{align}
	\expval*{V_{\text{int},\psi^*}} &=  \sum^{\infty}_{n=1} \alpha_{2n} \dfrac{2n}{(\sqrt{2m})^{2n}} \binom{2n-1}{n}   \psi_s^{n} \psi_s^{* \ n-1}  
 \nonumber 	\\
&	= \pdv{}{\psi_s^*} \left( \sum^{\infty}_{n=1} \alpha_{2n}  \binom{2n}{n}   \left(\dfrac{\abs{\psi_s}}{\sqrt{2m}}\right)^{2n} \right) \nonumber
\\ &	= {\mathcal{V}_{\text{int}}}_{,\psi_s^*}\, . \hspace{-.8cm}
\end{align}
 
A similar proof exists for a non-analytic potential. In this case, we assume that the potential is a function of $\phi^2$. Start with the identity
\begin{equation}
	V_{\text{int},\psi^*} = 2\phi \dfrac{e^{imt}}{\sqrt{2m}} V_{\text{int},\phi^2} \, .
	\label{NA-DV-Ex}
\end{equation}
Expanding $V_{\text{int}, \phi^2}$ around the condensate yields
\begin{equation}
	\label{V'}
	V_{\text{int},\psi^*} = \dfrac{1}{m} \left( \psi + \psi^* e^{2imt} \right) \sum_{n=0}^{\infty} \tilde \alpha_{n+1}(|\psi|^2)  (n+1) \left( \frac{Y}{m} \right)^{n} |\psi|^{2n} \, ,
\end{equation}
where
\begin{equation}
	\left( \dfrac{Y}{m} \right)^n =\dfrac{1}{(2m)^n} \sum_{k=0}^n \binom{n}{k} \left( \frac{\psi}{\psi^*} \right)^{2k-n} e^{-2imt(2k-n)} \, ,
\end{equation}
and recall that $\tilde \alpha_{n}=\frac{1}{n!}\frac{\partial^n V_{\text{int}}}{\partial \phi^{2n}}|_{\phi^2=|\psi|^2/m}$.
Note that in this expression, the expansion coefficients are field-dependent. We have
\begin{align}
	\expval{ V_{\text{int},\psi^*} } = &  \dfrac{1}{m}  \psi_s \sum_{n=1}^{\infty}  \tilde \alpha_{2n}(|\psi_s|^2)  (2n) \binom{2n-1}{n} \left( \dfrac{|\psi_s|^2}{2m} \right)^{2n-1} \nonumber \\
	& + \dfrac{1}{m}  \psi_s \sum_{n=0}^{\infty} \tilde \alpha_{2n+1}(|\psi_s|^2)  (2n+1)\binom{2n}{n} \left(\dfrac{|\psi_s|^2}{2m}\right)^{2n}  + \mathcal{O}(\epsilon^2)   \, .
\end{align}
It is easy to show that
\begin{equation}
	\label{alphas}
	(2n+1)  \tilde \alpha_{2n+1}(|\psi_s|^2) = m \pdv{\tilde{\alpha}_{2n}}{\abs{\psi_s}^2} \, ,
\end{equation}
which, after some simple manipulations yield
\begin{equation}
	\expval{ V_{\text{int},\psi^*} } = {\mathcal{V}_{\text{int}}}_{,\psi_s^*} + \mathcal{O}(\epsilon^2)\, ,
\end{equation}
as desired.

\subsection{Eq.~\eqref{V2-commute}: $(V_{\text{int},\psi^*})_2 = \psi_s^* \, {\mathcal{V}'_{\text{int}}} + \mathcal{O}(\epsilon^2)$}  
This relation may be proven by a similar procedure. 
 For analytic potentials, we have 
\begin{equation}
	(V_{\text{int},\psi^*})_2 =  \psi_s^* \sum^{\infty}_{n=1}\alpha_{2n} \dfrac{1}{({2m})^{n-1}}  \binom{2n-1}{n} (2n) \left(\abs{\psi_s}^2 \right)^{n}   + \mathcal{O}(\epsilon^2) \, ,
\end{equation}
which can be rearranged as
\begin{align}
	(V_{\text{int},\psi^*})_2 & = \psi_s^* \pdv{}{\abs{\psi_s}^2} \left( \sum^{\infty}_{n=1} \alpha_{2n}   \binom{2n}{n}\left( \dfrac{\abs{\psi_s}^2}{{2m}}\right)^{n}   \right) + \mathcal{O}(\epsilon^2) 
	\\ \nonumber & = \psi_s^* \, {\mathcal{V}'_{\text{int}}}+ \mathcal{O}(\epsilon^2)\, .
\end{align}

Likewise, for non-analytic potentials, using Eq.~\eqref{V'} we have 
\begin{align}
	\left( V_{\text{int},\psi^*} \right)_2 = & 
	\dfrac{1}{m} \psi^*_s \sum_{n=1}^{\infty} \tilde \alpha_{2n}(|\psi_s|^2 ) (2n) \binom{2n-1}{n}  \left( \frac{\abs{\psi_s}^2}{2m} \right)^{2n-1}   \nonumber \\
	& + \dfrac{1}{m} \psi^*_s\sum_{n=1}^{\infty} \tilde \alpha_{2n+1}(|\psi_s|^2) (2n+1)  \binom{2n}{n}   \left( \frac{\abs{\psi_s}^2}{2m} \right)^{2n} 
	+ \mathcal{O}(\epsilon^2) \, .
\end{align}
Using Eq.~\eqref{alphas} results in
\begin{eqnarray}
	(V_{\text{int},\psi^*})_2 = \psi^*_s \pdv{}{\abs{\psi_s}^2} \left( \sum^{\infty}_{n=1} \tilde{\alpha}_{2n}   \binom{2n}{n}\bigg( \dfrac{\abs{\psi_s}^{2}}{2m}\bigg)^{2n}   \right)  + \mathcal{O}(\epsilon^2) 
\nonumber 	\\
	= \psi_s^* \, {\mathcal{V}'_{\text{int}}} + \mathcal{O}(\epsilon^2) 	\, .
	\hspace{4.9cm}
\end{eqnarray}

\section{Analytic expression for $c_2$ (Eq.~\eqref{c1-c2})}
\label{App.c2}
The integrals in the definition of $c_2$ may be performed by the following procedure. We first evaluate the integral on $\tilde r'$, resulting in expressions involving the incomplete gamma function of the form $\Gamma(\nu,2 \tilde r^q)$ with $\nu=2/q$ or $3/q$. To carry out the second integral, we first expand the incomplete gamma function using
\ba 
\Gamma(\nu,x) = \Gamma(\nu) - x^\nu \sum_{k=0}^{\infty} \dfrac{(-x)^{k}}{(\nu+k)\, k!} \, .
\ea  
This expansion allows us to compute the remaining integral for each term. As the final step, the result can be resummed from which we obtain the following compact form in terms of the gamma and hypergeometric functions:
\ba 
c_2=\dfrac{2^{\frac{1}{q}-2}}{3 \Gamma (\frac{3}{q})^2} \, \left[ 6 \Gamma (\frac{2}{q}) \Gamma (\frac{3}{q})- q  \Gamma (\frac{5}{q}) \bigg(3 \, _2F_1(\frac{2}{q},\frac{5}{q};\frac{q+2}{q};-1)-2 \, _2F_1(\frac{3}{q},\frac{5}{q};\frac{q+3}{q};-1)\bigg) \right] \, .
\ea 
For $q=1$, this expression reduces to the well-known result, i.e., $c_2=\frac{5}{16}$ (see e.g., Ref.~\cite{Deng:2018}).


\end{document}